\documentclass{bmvc2k}

\title{
Learning Event-guided Exposure-agnostic Video Frame Interpolation via Adaptive Feature Blending
}

\addauthor{Junsik Jung}{junsik.jung@kaist.ac.kr}{1}
\addauthor{Yoonki Cho}{yoonki@kaist.ac.kr}{1}
\addauthor{Woo Jae Kim}{wkim97@kaist.ac.kr}{1}
\addauthor{Lin Wang}{linwang@ntu.edu.sg}{2}
\addauthor{Sung-eui Yoon}{sungeui@kaist.edu}{1}

\addinstitution{
 School of Computing,\\
 Korea Advanced Institute of Science and Technology (KAIST),\\
 Daejeon, Korea
}
\addinstitution{
 School of Electrical and Electronic Engineering,\\
 Nanyang Technological University (NTU),\\
 Singapore
}

\usepackage{amssymb}

\usepackage{booktabs}
\usepackage{amsmath}
\usepackage{bm}
\usepackage{wrapfig}
\usepackage[dvipsnames]{xcolor}
\usepackage{orcidlink}
\usepackage{pifont}
\usepackage{multirow}
\newcommand{\cmark}{\ding{51}}%
\newcommand{\xmark}{\ding{55}}%
\newcommand{\womark}{\ding{53}}%

\usepackage{colortbl}
\definecolor{tableHeadGray}{gray}{.9}
\definecolor{tableSubHeadGray}{gray}{0.95}

\newcommand{\Skip}[1]{}

\runninghead{Jung \etal}{Exposure-agnostic VFI via Adaptive Feature Blending}

\def\etal{\emph{et al}\bmvaOneDot}

\begin{document}

\maketitle

\begin{abstract}
Exposure-agnostic video frame interpolation (VFI) is a challenging task that aims to recover sharp, high-frame-rate videos from blurry, low-frame-rate inputs captured under unknown and dynamic exposure conditions.
Event cameras are sensors with high temporal resolution, making them especially advantageous for this task.
However, existing event-guided methods struggle to produce satisfactory results on severely low-frame-rate blurry videos due to the lack of temporal constraints.
In this paper, we introduce a novel event-guided framework for exposure-agnostic VFI, addressing this limitation through two key components: a Target-adaptive Event Sampling (TES) and a Target-adaptive Importance Mapping (TIM).  
Specifically, TES samples events around the target timestamp and the unknown exposure time to better align them with the corresponding blurry frames.
TIM then generates an importance map that considers the temporal proximity and spatial relevance of consecutive features to the target.
Guided by this map, our framework adaptively blends consecutive features, allowing temporally aligned features to serve as the primary cues while spatially relevant ones offer complementary support. Extensive experiments on both synthetic and real-world datasets demonstrate the effectiveness of our approach in exposure-agnostic VFI scenarios.
\end{abstract}

\vspace{-3mm}
\section{Introduction}
\label{sec:intro}

Video frame interpolation (VFI) aims to generate intermediate frames between consecutive inputs, converting low-frame-rate videos into high-frame-rate ones~\cite{interp_survey}. Recent advances in event cameras~\cite{event_vision_survey}, which asynchronously capture per-pixel brightness changes with high temporal resolution, have significantly improved VFI performance~\cite{timelens, timereplayer, timelenspp, ledvdi, evdi, refid}, particularly in exposure-specific settings where the camera exposure time is fixed and known.
However, in real-world scenarios, exposure time often varies and is blind due to auto-exposure mechanisms~\cite{cinematography}, resulting in inputs with inconsistent motion blur. This variability poses challenges for conventional VFI models, which are typically designed for either consistently sharp or blurry inputs. Consequently, there is a growing need for exposure-agnostic VFI, which can handle unknown and dynamic exposure conditions and is thus essential for practicality.

\begin{figure}[t]
    \centering
    \subfigure[]{%
        \includegraphics[width=0.22\textwidth]{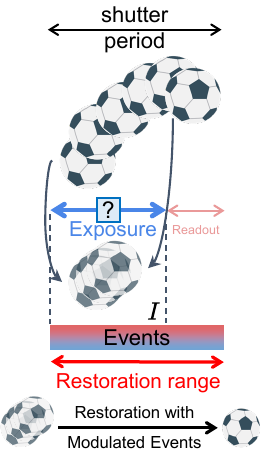}
        \label{fig:intro_sub1-2}
    }
    \hfill
    \subfigure[]{%
        \includegraphics[width=0.32\textwidth]{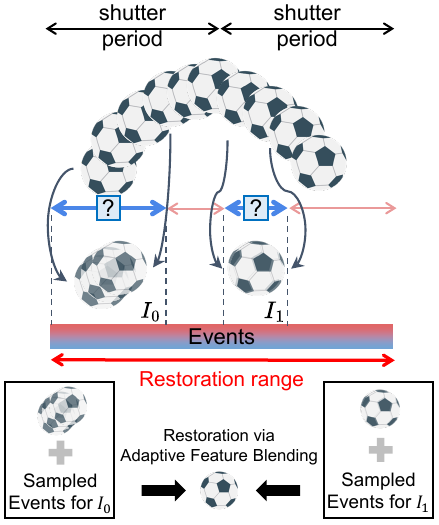}
        \label{fig:intro_sub2}
    }
    \hfill
    \subfigure[]{%
        \includegraphics[width=0.30\textwidth]{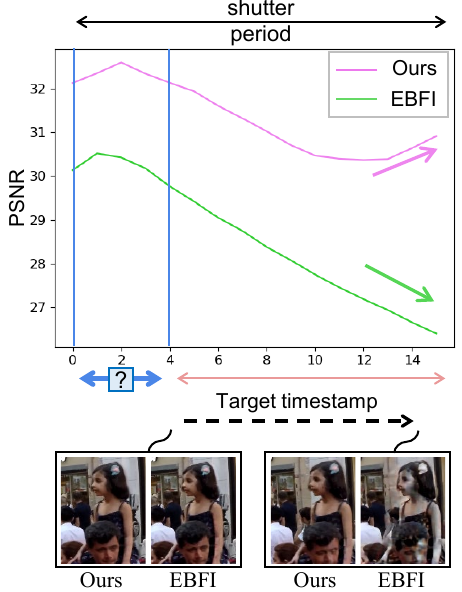}
        \label{fig:intro_sub3}
    }
    \caption{
        Schematic comparison with a prior event-guided exposure-agnostic VFI method.
        (a) EBFI~\cite{EBFI} restores sharp frames from a single blurry input via event modulation, but degrades as the target timestamp deviates due to the lack of temporal constraints.
        (b) Our method addresses this by explicitly leveraging temporal constraints via adaptive feature blending.
        (c) PSNR comparison on GoPro~\cite{gopro} shows that our method maintains more stable performance across varying timestamps.
    }
    \vspace{-5.3mm}
    \label{fig:intro}
\end{figure}

Recently, a few event-guided methods~\cite{uevd, EBFI} have been proposed for exposure-agnostic video restoration.
UEVD~\cite{uevd} introduced an event-guided method for video deblurring under blind exposure conditions.
However, it is limited to restoring a single sharp frame within the exposure window corresponding to the blurry input, and requires a separate VFI step to generate high-frame-rate outputs.
EBFI~\cite{EBFI} later proposed a unified framework that restores arbitrary target frames under blind exposure from a single blurry frame by modulating events (see Fig.~\ref{fig:intro_sub1-2}). Yet, the lack of temporal constraints — that is, leveraging the temporal relationships between neighboring frames for consistent and accurate guidance — leads to performance degradation, especially when input videos are captured at very low frame rates.


To address this limitation, a promising direction is to introduce \textit{temporal constraints across consecutive frames} since frames closer to the target timestamp often offer more relevant cues.
Flow-based methods~\cite{rife, ifrnet, timelens, timelenspp, timereplayer} offer a natural way to model such constraints by aligning frame-to-frame correspondences. However, their effectiveness is limited under blind exposure, where severe motion blur degrades both frame quality and flow estimation~\cite{interp_survey, event_vision_survey}.
An alternative is to leverage the \textit{mutual relationship} between frames and events~\cite{edi}, which introduces two key challenges:
(1) how to sample events between the target timestamp and the unknown exposure period in a way that reflects the mutual relationship between frames and events, and
(2) how to impose temporal constraints across consecutive features while jointly considering temporal proximity and spatial relevance.


In this paper, we address the aforementioned challenges through a simple yet effective adaptive feature blending strategy (see Fig.~\ref{fig:intro_sub2}).
To this end, we introduce a novel learning framework that consists of two key modules: the Target-adaptive Event Sampling (TES) and the Target-adaptive Importance Mapping (TIM).
TES samples events around the target timestamp and the unknown exposure period, which are fused with the corresponding blurry frames to guide target sharp frame synthesis.
TIM is designed to generate an importance map that adaptively weights features from consecutive frames based on their temporal proximity and spatial relevance to the target.
This allows our model to incorporate temporal constraints and maintain performance even as the target timestamp deviates (see Fig.\ref{fig:intro_sub3}).
Experiments on synthetic and real-world datasets show that our method consistently outperforms existing approaches under blind exposure, achieving at least a 2dB PSNR improvement.

In summary, our contributions are threefold:
1) We propose a simple yet effective feature blending strategy that, for the first time, incorporates temporal constraints into event-guided VFI under unknown and dynamic exposure conditions.
2) We design a novel unified framework with two synergistic modules: TES, which samples events around the target timestamp and unknown exposure; and TIM, which generates an importance map to adaptively weight features based on temporal and spatial relevance.
3) We validate our approach through extensive experiments on synthetic and real-world datasets, showing its effectiveness across diverse blind exposure scenarios.

\begin{figure*}[t]
    \centering
    \includegraphics[clip, width=0.95\textwidth]{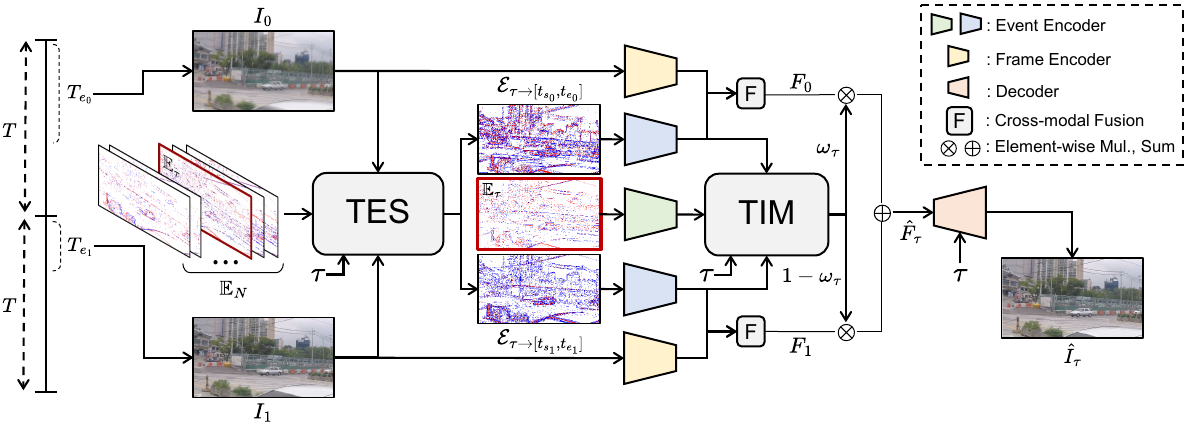}    
    \caption{
        Overview of our framework.
        Given blurry frames $(I_0, I_1)$ with unknown exposures $(T_{e_0}, T_{e_1})$, stacked events $\mathbb{E}_N$ over $2T$, and target timestamp $\tau$, the model reconstructs the target frame $\hat{I}_\tau$.
        TES samples events around $\tau$ and unknown exposure, which are fused with frames.
        TIM generates an importance map $\omega_\tau$ to adaptively blend the fused features $(F_0, F_1)$ into $F_\tau$, which is then decoded into $\hat{I}_\tau$.
        Shared encoders are color-coded.
    }
    \label{fig:overview}
\end{figure*}

\vspace{-5mm}
\section{Related Work}
\vspace{-3mm}

\noindent \textbf{Exposure-specific Video Restoration.} 
Most prior work assumes a known and fixed exposure time when restoring sharp, high-frame-rate videos.
For video frame interpolation (VFI), existing methods are broadly categorized into flow-based~\cite{longterm, soft_splatting, rife, ifrnet, many_splatting, film}, kernel-based~\cite{adacof, adaconv}, architecture-based~\cite{vfi_trans1, vfi_trans2, cain, GDconvnet}, and event-guided~\cite{timelens, timelenspp, timereplayer} approaches.
Motion deblurring has also been studied independently to restore one sharp frame from each blurry input~\cite{gatedattention, multi, edi, lebmd, efnet, ResidualDeblur, nafnet}.
Rather than treating VFI and motion deblurring as separate tasks, recent studies have explored unified frameworks that address both jointly, using either RGB-only~\cite{deblurnet_interpnet, pyramid} or event-guided settings~\cite{ledvdi, evdi, refid}.
However, their reliance on fixed exposure limits their applicability to real-world scenarios where exposure is unknown and dynamically changing.
\noindent \textbf{Exposure-agnostic Video Restoration.} 
A few RGB-only methods~\cite{prior, vidue} have been proposed for video restoration under unknown exposure conditions. However, due to the lack of precise motion cues, these approaches suffer from motion ambiguity~\cite{EBFI, refid}, making it difficult to accurately recover sharp frames. To overcome this, event cameras offer a promising alternative by providing temporally dense and precise motion information.
Building on this idea, Kim~\etal~\cite{uevd} proposed an event-guided motion deblurring method under blind exposure. However, their approach is limited to reconstructing a single sharp frame within the exposure window and requires an additional VFI step to generate high-frame-rate output.
Weng~\etal~\cite{EBFI} presented a framework for interpolating arbitrary frames by modulating a blurry frame with events, but the lack of temporal constraints leads to degraded performance under severe blur and sparse temporal sampling.
We address this limitation with a unified framework that explicitly incorporates temporal constraints into event-guided, exposure-agnostic VFI.

\begin{figure}[t]
    \centering
    \subfigure[TES module]{%
        \includegraphics[width=0.49\linewidth]{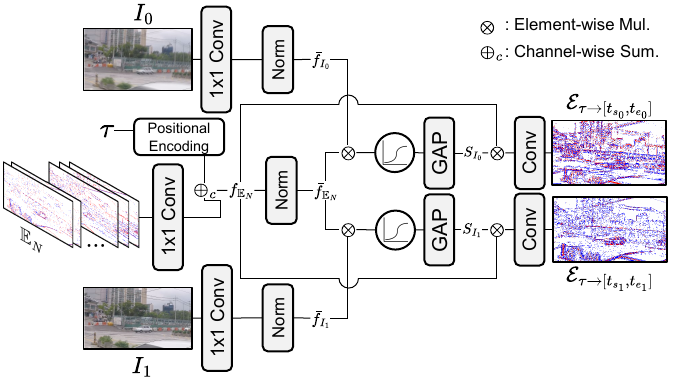}
        \label{fig:tes}
    }
    \hfill
    \subfigure[TIM module]{%
        \includegraphics[width=0.49\linewidth]{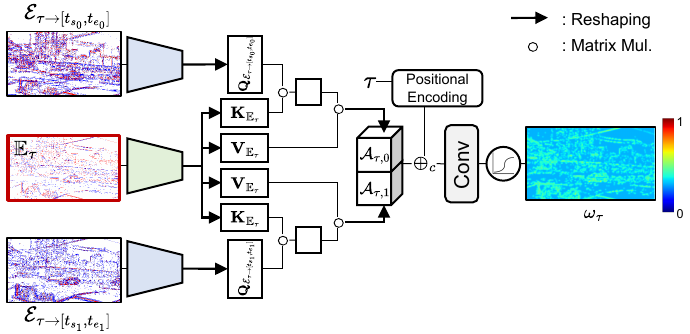}
        \label{fig:tim}
    }
    \caption{Architectures of the TES (Sec.~\ref{ssec:tes}) and TIM (Sec.~\ref{ssec:tim}) modules.}
    \vspace{-5mm}
\end{figure}

\vspace{-4mm}
\section{Method}
\label{sec:method}
\vspace{-3mm}
In this section, we first formulate the problem based on the mutual relationship between frames and events using the Event-based Double Integral (EDI) model~\cite{edi}.
We then present an overview of our framework (Sec.\ref{ssec:overview}), followed by details of the proposed TES and TIM modules (Sec.\ref{ssec:tes},~\ref{ssec:tim}).

\noindent \textbf{Problem Formulation.}
Event cameras capture events, denoted as $e(\mathcal{T})$, at timestamp $\mathcal{T}$ across the pixel space asynchronously~\cite{event_vision_survey}. 
These events are recorded when the logarithmic change in intensity exceeds a contrast threshold $c$.
During the exposure time $T_e$, the captured frame $I$ is derived by averaging latent frames $L(t)$ present in $T_e$ as:
\begin{equation}
    I = \frac{1}{T_e}\int_{t_s}^{t_e} L(t) dt,
\label{eq:captured_frame}
\end{equation}
where $L$, $t_s$, and $t_e$ denote the latent frame, the starting timestamp of $T_e$, and the end timestamp of $T_e$, respectively.

Given latent frames $L(\tau)$ and $L(t)$ at different timestamps $\tau$ and $t$ during the exposure time, and the events spanning the interval $[\tau, t]$, we can establish the following relationship between the latent frames:
\begin{equation}
\begin{split}
    L(t) & = L(\tau) \exp \left( c\cdot \mathcal{E}_{\tau \rightarrow t} \right),
\end{split}
\label{eq:frame_event_relationship}
\end{equation}
where $\mathcal{E}_{\tau \rightarrow t}=\int_{\tau}^t e(x) dx$ denotes a 2D tensor obtained by stacking events from $\tau$ to $t$.
Combining Eqs. (\ref{eq:captured_frame}) and (\ref{eq:frame_event_relationship}), the following Event-based Double Integral (EDI) model~\cite{edi} can be derived:
\begin{equation}
L(\tau) = \frac{T_e \cdot I}{\int_{t_s}^{t_e} \exp \left(c\cdot \mathcal{E}_{\tau \rightarrow t} \right) dt}.
\label{eq:EDI}
\end{equation}

The EDI model describes how the latent frame $L(\tau)$ can be reconstructed by combining the captured frame $I$ with stacked events over the exposure duration $T_e$.
In VFI scenarios where the target timestamp $\tau$ lies beyond $T_e$, EVDI~\cite{evdi} extended such mutual relationship by using events from $\tau$ to $T_e$ to restore sharp frames outside the exposure window.
However, this approach requires precise knowledge of $T_e$, limiting its practicality.
Nonetheless, this suggests that if we can approximate events centered around $\tau$ and $T_e$ without knowing $T_e$, it is still possible to recover $L(\tau)$ based on Eq.~(\ref{eq:EDI}).
We denote this event set as $\mathcal{E}_{\tau \rightarrow [t_s, t_e]}$.


\vspace{-4mm}
\subsection{Overview}
\label{ssec:overview}
\vspace{-2mm}
We adopt a UNet-based encoder-decoder model~\cite{Unet}, comprising three encoders and two decoders, with a refinement block~\cite{evdi} appended at the final decoder stage.
The model takes as input two consecutive captured frames $(I_0, I_1)$ — each recorded under unknown exposure durations $(T_{e_0}, T_{e_1})$ — along with the corresponding events recorded during their shutter periods $2T$.
These events are converted into event frames~\cite{event_driving} by pixel-wise accumulation over $N$ bins, yielding $\mathbb{E}_N \in \mathbb{R}^{N \times H \times W}$, where $N$ denotes the number of temporal slices.

To apply Eq.~(\ref{eq:EDI}), we sample a subset of events $\mathcal{E}_{\tau \rightarrow [t_s, t_e]}$ from the stacked input $\mathbb{E}_N$, centered around the target timestamp $\tau$ and unknown exposure $T_e$.
To this end, we introduce the Target-adaptive Event Sampling (TES) module (Sec.~\ref{ssec:tes}), samples $(\mathcal{E}_{\tau \rightarrow [t_{s_0}, t_{e_0}]}, \mathcal{E}_{\tau \rightarrow [t_{s_1}, t_{e_1}]})$ accordingly.
Frame and event features from $(I_0, I_1)$ and sampled events are fused via a cross-modal technique~\cite{efnet} to produce $(F_0, F_1)$.

To leverage temporal constraints, we introduce the Target-adaptive Importance Mapping (TIM) module (Sec.~\ref{ssec:tim}), which generates an importance map $\omega_\tau$ based on the temporal proximity and spatial relevance of $(F_0, F_1)$ to $\tau$.
This map guides adaptive blending: temporally closer features are emphasized, while spatially relevant cues complement them.
The resulting fused feature $\hat{F}_\tau$ is decoded to reconstruct the target frame $\hat{I}_\tau$.

\vspace{-3mm}
\subsection{Target-adaptive Event Sampling}
\label{ssec:tes}
Recalling Eq.~(\ref{eq:EDI}), the frames $(I_0, I_1)$ should be combined with events centered around the target timestamp $\tau$ and the unknown exposure durations $(T_{e_0}, T_{e_1})$.
To achieve this, we follow prior methods~\cite{EBFI, uevd} by using the captured frames as references under blind exposure.
However, unlike previous approaches that focus solely on the unknown exposure period, our module is additionally designed to account for the target timestamp $\tau$.

A detailed illustration of the TES module is shown in Fig.~\ref{fig:tes}.
For clarity, we describe the process of sampling the stacked events for $I_0$, denoted as $\mathcal{E}_{\tau \rightarrow [t_{s_0}, t_{e_0}]}$.
Given $\mathbb{E}_{N}$, $I_0$, and $\tau$ as inputs, the TES module is formulated as: 
\begin{equation} 
\mathcal{E}_{\tau \rightarrow [t_{s_0}, t_{e_0}]} = \text{TES}(\mathbb{E}_{N}, I_0, \tau). 
\label{eq:EES module} 
\end{equation}


We first extract features from $\mathbb{E}_N$ and $I_0$.
To incorporate the target timestamp $\tau$, we add its positional encoding to the event features channel-wise, yielding $f_{\mathbb{E}_N}$.
We then compute the correlation score $\mathcal{S}_{I_0}$ between the normalized features of $\bar{f}_{I_0}$ and $\bar{f}_{\mathbb{E}_N}$ as:
\begin{equation} 
\mathcal{S}_{I_0} = \text{GAP}\big( \sigma(\bar{f}_{I_0} \otimes \bar{f}_{\mathbb{E}_N}) \big), 
\label{eq:corr_score} 
\end{equation} 
where $\otimes$, $\sigma$, and GAP denote element-wise multiplication, the sigmoid function, and global average pooling, respectively.
Using $\mathcal{S}_{I_0}$, we perform element-wise multiplication with $f_{\mathbb{E}_N}$ to sample the event features centered around $\tau$ and $T_{e_0}$.
A convolution is then applied to produce the final representation $\mathcal{E}_{\tau \rightarrow [t_{s_0}, t_{e_0}]}$.
The counterpart $\mathcal{E}_{\tau \rightarrow [t_{s_1}, t_{e_1}]}$ is derived analogously using $I_1$, and both event representations are fused with $(I_0, I_1)$ using the fusion method from~\cite{efnet} (denoted as `F' in Fig.~\ref{fig:overview}).

\subsection{Target-adaptive Importance Mapping}
\label{ssec:tim}
\vspace{-1mm}
Although TES provides relevant events, restoration degrades as $\tau$ shifts from the exposure due to event sensor noise~\cite{event_vision_survey}. 
When $\tau$ nears $T_{e_1}$, relying only on $I_0$ becomes suboptimal, making features from $I_1$ more informative and underscoring the need for temporal constraints.
To address this, we introduce the TIM module, which estimates an importance map $\omega_\tau$ to adaptively weight fused features $(F_0, F_1)$ based on their temporal proximity and spatial relevance to the target.
Given the sampled events $(\mathcal{E}_{\tau \rightarrow [t_{s_0}, t_{e_0}]}, \mathcal{E}_{\tau \rightarrow [t_{s_1}, t_{e_1}]})$, the $\tau$-centered event stack $\mathbb{E}_\tau$, and the timestamp $\tau$, TIM estimates $\omega_\tau$ as:
\begin{equation}
\omega_\tau = \text{TIM} ( \mathcal{E}_{\tau \rightarrow [t_{s_0}, t_{e_0}]}, \mathcal{E}_{\tau \rightarrow [t_{s_1}, t_{e_1}]}, \mathbb{E}_\tau, \tau ).
\label{eq:TIM}
\end{equation}

A detailed illustration of the TIM module is shown in Fig.\ref{fig:tim}.
We first extract features from the sampled events $(\mathcal{E}_{\tau \rightarrow [t_{s_0}, t_{e_0}]}, \mathcal{E}_{\tau \rightarrow [t_{s_1}, t_{e_1}]})$ and the $\tau$-centered event stack $\mathbb{E}_\tau$.
To estimate spatial relevance efficiently, we apply channel-wise attention~\cite{RUN}: features from the sampled events are used as queries, and those from $\mathbb{E}_\tau$ serve as keys and values.
After flattening the spatial dimensions, the attended feature $\mathcal{A}_{\tau,0}$ is computed as:
\begin{equation}
    \mathcal{A}_{\tau,0} = \mathbf{V}_{\mathbb{E}_\tau} \circ \text{softmax}\left(\frac{{\mathbf{Q}_{\mathcal{E}_{\tau \rightarrow [t_{s_0}, t_{e_0}]}}}^T \circ \mathbf{K}_{\mathbb{E}_\tau}}{\sqrt{d_k}}\right),
\label{eq:tim_step1}
\end{equation}
where $d_k$ denotes the key dimension.
$\mathcal{A}_{\tau,1}$ is computed analogously. The two attended features are concatenated and combined with the positional encoding of $\tau$ via channel-wise addition to encode temporal information.
The result is passed through convolutional layers and a sigmoid activation to produce the importance map $\omega_\tau$, which encodes both temporal proximity and spatial relevance.

The target feature $\hat{F}_\tau$ is obtained via an adaptive blending strategy that applies temporal constraints:
\begin{equation}
   \hat{F}_\tau = \big( \omega_\tau \otimes F_0 \big) \oplus \big( (1-\omega_\tau) \otimes F_1 \big).
   \label{eq:TFB}
\end{equation}
A decoder then transforms $\hat{F}_\tau$ into the target sharp frame $\hat{I}_\tau$.
This blending strategy enables the model to effectively leverage temporal constraints between consecutive features.
The entire framework is trained end-to-end using the Charbonnier loss~\cite{char}:
\begin{equation}
    \mathcal{L} = \sqrt{ \|I^{gt}_\tau - \hat{I}_\tau\|^2 + \epsilon^2}, \text{with $\epsilon=$\texttt{1e-6}.}
\label{eq:charloss}
\end{equation}

\begin{table}[t] 
\caption{
Summary of compared methods.
‘Frame’ and ‘Events’ denote input types; ‘Agnostic’ and ‘Unified’ indicate support for blind exposure and joint deblurring-interpolation; ‘T-Constraint’ refers to the use of temporal constraints.
}
\centering
\small
\resizebox{0.63\textwidth}{!}{
\setlength{\tabcolsep}{10pt}
\renewcommand\arraystretch{1.0}
    \begin{tabular}{ l c c  c  c c c}
        \bottomrule[0.15em]
        \rowcolor{tableHeadGray}
        \textbf{Method}                     &                & \textbf{Frame} & \textbf{Events} & \textbf{Agnostic} & \textbf{Unified} & \textbf{T-Constraint}
        \\ \hline \hline
        \multicolumn{2}{l}{NAFNet~\cite{nafnet}+RIFE~\cite{rife}}            & \cmark   & \xmark & \xmark & \xmark & \cmark (via RIFE)\\
        \multicolumn{2}{l}{UTI~\cite{prior}}              & \cmark   & \xmark & \cmark & \cmark & \cmark \\
        \multicolumn{2}{l}{VIDUE~\cite{vidue}}            & \cmark   & \xmark & \cmark & \cmark & \cmark\\
        \multicolumn{2}{l}{UEVD~\cite{uevd}+TL~\cite{timelens}}  & \cmark   & \cmark & \cmark (via UEVD) & \xmark & \cmark (via TL) \\ 
        \multicolumn{2}{l}{EVDI~\cite{evdi}}                 & \cmark   & \cmark  & \xmark & \cmark & \cmark\\ 
        \multicolumn{2}{l}{REFID~\cite{refid}}                 & \cmark   & \cmark  & \xmark & \cmark & \cmark\\ 
        \multicolumn{2}{l}{EBFI~\cite{EBFI}}                 & \cmark   & \cmark  & \cmark & \cmark & \xmark \\  \hline
        \multicolumn{2}{l}{Ours}                 & \cmark   & \cmark  & \cmark & \cmark & \cmark \\ \hline
    \end{tabular}
    }
\vspace{-3mm}
\label{tab:summary}
\end{table}

\begin{table*}[t]
\centering
\small
\setlength{\tabcolsep}{8pt}
\caption{Quantitative results on GoPro and HighREV (10$\Downarrow$). \textbf{Bold} and \underline{underlined} indicate the best and second-best scores, respectively.}
\resizebox{0.89\linewidth}{!}
        {
        \begin{tabular}{lcccccccc}
        \bottomrule[0.15em]
        \rowcolor{tableHeadGray}
        \textbf{Method}       &  \multicolumn{8}{c}{\textbf{PSNR $\uparrow$ / SSIM $\uparrow$ / LPIPS $\downarrow$}}     \\ \hline
         \hline
        \textit{GoPro}-10$\Downarrow$            & \multicolumn{2}{c}{\textbf{9+1}}   & \multicolumn{2}{c}{\textbf{5+5}}  &  \multicolumn{2}{c}{\textbf{1+9}} & \multicolumn{2}{c}{\textbf{RandEx}} \\ \hline
        NAFnet~\cite{nafnet}+RIFE~\cite{rife}     &  \multicolumn{2}{c}{23.15 / 0.736 / 0.164}        & \multicolumn{2}{c}{21.82 / 0.665 / 0.188} & \multicolumn{2}{c}{18.78 / 0.520 / 0.282} & \multicolumn{2}{c}{21.05 / 0.633 / 0.219}\\
        UTI~\cite{prior}   &   \multicolumn{2}{c}{23.84 / 0.805 / 0.205}    & \multicolumn{2}{c}{23.25 / 0.771 / 0.166} & \multicolumn{2}{c}{20.95 / 0.640 /0.170} & \multicolumn{2}{c}{22.17 / 0.714 / 0.189} \\
        VIDUE~\cite{vidue}     &  \multicolumn{2}{c}{25.37 / 0.818 / 0.181}        & \multicolumn{2}{c}{26.24 / 0.835 / 0.145} & \multicolumn{2}{c}{24.76 / 0.764 / 0.146} & \multicolumn{2}{c}{25.86 / 0.819 / 0.152}\\
        UEVD~\cite{uevd}+TL~\cite{timelens}  &   \multicolumn{2}{c}{23.40 / 0.720 / 0.194}    & \multicolumn{2}{c}{23.31 / 0.755 / 0.161} & \multicolumn{2}{c}{26.33 / 0.866 / 0.103} & \multicolumn{2}{c}{23.51 / 0.757 / 0.164} \\
        EVDI~\cite{evdi}        &   \multicolumn{2}{c}{28.18 / 0.898 / \underline{0.052}}          & \multicolumn{2}{c}{27.73 / 0.884 / \underline{0.067}}          & \multicolumn{2}{c}{26.07 / 0.834 / 0.084}          & \multicolumn{2}{c}{27.51 / 0.877 / \underline{0.071}}\\
        REFID~\cite{refid}      & \multicolumn{2}{c}{\underline{31.08} / 0.939 / 0.077}          & \multicolumn{2}{c}{\underline{31.28} / \underline{0.941} / 0.071}          & \multicolumn{2}{c}{30.27 / \underline{0.927} / \underline{0.078}}          & \multicolumn{2}{c}{\underline{31.03} / \underline{0.938} / 0.074}\\
        EBFI~\cite{EBFI} &   \multicolumn{2}{c}{31.06 / \underline{0.942} / 0.072}          & \multicolumn{2}{c}{31.08 / 0.940 / 0.071}          & \multicolumn{2}{c}{\underline{30.32} / 0.921 / 0.087}          & \multicolumn{2}{c}{30.89 / 0.936 / 0.075}\\
        \textbf{Ours}       &   \multicolumn{2}{c}{\textbf{33.22 / 0.960 / 0.050}} & \multicolumn{2}{c}{\textbf{33.61 / 0.963 / 0.042}} & \multicolumn{2}{c}{\textbf{32.87 / 0.954 / 0.048}} & \multicolumn{2}{c}{\textbf{33.39 / 0.961 / 0.045}}\\ \hline
        \textit{HighREV}-10$\Downarrow$ &          &                                                       &                                                      &                                    \\ \hline
        NAFnet~\cite{nafnet}+RIFE~\cite{rife}     &  \multicolumn{2}{c}{25.09 / 0.805 / 0.446}        & \multicolumn{2}{c}{24.90 / 0.874 / 0.401} & \multicolumn{2}{c}{26.07 / 0.897 / 0.361} & \multicolumn{2}{c}{25.20 / 0.844 / 0.402}\\
        UTI~\cite{prior}  &   \multicolumn{2}{c}{26.22 / 0.834 / 0.371}    & \multicolumn{2}{c}{26.15 / 0.821 / 0.395} & \multicolumn{2}{c}{26.51 / 0.851 / 0.395} & \multicolumn{2}{c}{26.19 / 0.831 / 0.398} \\
        VIDUE~\cite{vidue}     &  \multicolumn{2}{c}{26.65 / 0.847 / 0.375}                              & \multicolumn{2}{c}{27.43 / 0.860 / 0.367} & \multicolumn{2}{c}{25.31 / 0.841 / 0.387} & \multicolumn{2}{c}{26.76 / 0.850 / 0.373}\\
        UEVD~\cite{uevd}+TL~\cite{timelens}  &  \multicolumn{2}{c}{26.02 /0.844 / 0.399}                              & \multicolumn{2}{c}{27.41 / 0.854 / 0.378} & \multicolumn{2}{c}{27.77 / 0.861 / 0.337} & \multicolumn{2}{c}{27.43 / 0.854 / 0.373} \\
        EVDI~\cite{evdi}   &  \multicolumn{2}{c}{30.26 / 0.910 / \underline{0.278}}     & \multicolumn{2}{c}{29.75 / 0.896 / \underline{0.277}} & \multicolumn{2}{c}{27.01 / 0.839 / 0.292} & \multicolumn{2}{c}{29.25 / 0.887 / \underline{0.273}}\\
        REFID~\cite{refid}   &  \multicolumn{2}{c}{\underline{33.60} / \underline{0.937} / 0.306}   & \multicolumn{2}{c}{\underline{34.18} / \underline{0.938} / 0.296} & \multicolumn{2}{c}{\underline{33.38} / \underline{0.928} / \underline{0.260}} & \multicolumn{2}{c}{\underline{33.91} / \underline{0.936} / 0.292}\\
        EBFI~\cite{EBFI}   &  \multicolumn{2}{c}{28.23 / 0.898 / 0.362}          & \multicolumn{2}{c}{27.84 / 0.887 / 0.342} & \multicolumn{2}{c}{26.25 / 0.855 / 0.279} & \multicolumn{2}{c}{27.55 / 0.882 / 0.333}\\
        \textbf{Ours}          &  \multicolumn{2}{c}{\textbf{35.73 / 0.948 / 0.263}} & \multicolumn{2}{c}{\textbf{36.27 / 0.949 / 0.250}} & \multicolumn{2}{c}{\textbf{35.45 / 0.942 / 0.248}} & \multicolumn{2}{c}{\textbf{36.02 / 0.947 / 0.253}}\\ \hline
        \end{tabular}
        }
\vspace{-4mm}
\label{tab:gopro_highrev_10sp}
\end{table*}

\begin{table*}[!t]
\centering
\small
\setlength{\tabcolsep}{5pt}
\caption{Quantitative results on GoPro and HighREV (16$\Downarrow$).}
\resizebox{0.89\textwidth}{!}
        {
        \begin{tabular}{lcccccccccc}
        \bottomrule[0.15em]
        \rowcolor{tableHeadGray}
        \textbf{Method}         &  \multicolumn{10}{c}{\textbf{PSNR $\uparrow$ / SSIM $\uparrow$ / LPIPS $\downarrow$}}     \\ \hline
         \hline
        \textit{GoPro}-16$\Downarrow$  & \multicolumn{2}{c}{\textbf{15+1}}                    & \multicolumn{2}{c}{\textbf{11+5}} & \multicolumn{2}{c}{\textbf{5+11}} & \multicolumn{2}{c}{\textbf{1+15}} & \multicolumn{2}{c}{\textbf{RandEx}}   \\ \hline
        NAFnet~\cite{nafnet}+RIFE~\cite{rife}      & \multicolumn{2}{c}{21.37 / 0.645 / 0.297}                 & \multicolumn{2}{c}{21.10 / 0.627 / 0.284} & \multicolumn{2}{c}{19.75 / 0.551 / 0.270} & \multicolumn{2}{c}{17.32 / 0.430 / 0.322} & \multicolumn{2}{c}{20.03 / 0.569 / 0.233} \\
        UTI~\cite{prior}   & \multicolumn{2}{c}{21.43 / 0.672 / 0.286} & \multicolumn{2}{c}{21.19 / 0.659 / 0.273} & \multicolumn{2}{c}{20.23 / 0.591 / 0.272} & \multicolumn{2}{c}{19.64 / 0.534 / 0.298} & \multicolumn{2}{c}{20.74 / 0.609 / 0.234} \\
        VIDUE~\cite{vidue}      & \multicolumn{2}{c}{22.44 / 0.686 / 0.298}                 & \multicolumn{2}{c}{23.27 / 0.721 / 0.270} & \multicolumn{2}{c}{23.10 / 0.697 / 0.231} & \multicolumn{2}{c}{21.58 / 0.623 / 0.220} & \multicolumn{2}{c}{22.89 / 0.695 / 0.253} \\
        UEVD~\cite{uevd}+TL~\cite{timelens}   & \multicolumn{2}{c}{17.04 / 0.316 / 0.472} & \multicolumn{2}{c}{17.26 / 0.334 / 0.459} & \multicolumn{2}{c}{17.92 / 0.376 / 0.444} & \multicolumn{2}{c}{18.41 / 0.401 / 0.438} & \multicolumn{2}{c}{17.64 / 0.357 / 0.453} \\
        EVDI~\cite{evdi}        & \multicolumn{2}{c}{27.38 / 0.881 / \underline{0.083}}          & \multicolumn{2}{c}{27.16 / 0.874 / \underline{0.084}} & \multicolumn{2}{c}{26.36 / 0.847 / \underline{0.092}} & \multicolumn{2}{c}{24.15 / 0.774 / \underline{0.115}} & \multicolumn{2}{c}{26.52 / 0.853 / \underline{0.088}} \\
        REFID~\cite{refid}  & \multicolumn{2}{c}{\underline{28.98} / 0.912 / 0.118}          & \multicolumn{2}{c}{\underline{29.11} / 0.913 / 0.115} & \multicolumn{2}{c}{\underline{28.84} / \underline{0.909} / 0.115} & \multicolumn{2}{c}{\underline{28.16} / \underline{0.896} / 0.123} & \multicolumn{2}{c}{\underline{28.87} / \underline{0.909} / 0.116} \\
        EBFI~\cite{EBFI} & \multicolumn{2}{c}{28.78 / \underline{0.913} / 0.108}          & \multicolumn{2}{c}{28.92 / \underline{0.914} / 0.106} & \multicolumn{2}{c}{28.48 / 0.903 / 0.116} & \multicolumn{2}{c}{27.64 / 0.883 / 0.134} & \multicolumn{2}{c}{28.59 / 0.906 / 0.113} \\
        \textbf{Ours} & \multicolumn{2}{c}{\textbf{30.97 / 0.937 / 0.081}} & \multicolumn{2}{c}{\textbf{31.39 / 0.941 / 0.074}} & \multicolumn{2}{c}{\textbf{31.18 / 0.938 / 0.075}} & \multicolumn{2}{c}{\textbf{30.07 / 0.925 / 0.085}} & \multicolumn{2}{c}{\textbf{31.14 / 0.938 / 0.076}} \\ \hline

        \textit{HighREV}-16$\Downarrow$          &                    &               &           &               &               \\ \hline
        NAFnet~\cite{nafnet}+RIFE~\cite{rife}      & \multicolumn{2}{c}{19.85 / 0.336 / 0.744}                 & \multicolumn{2}{c}{19.70 / 0.285 / 0.770} & \multicolumn{2}{c}{19.91 / 0.312 / 0.764} & \multicolumn{2}{c}{20.35 / 0.332 / 0.734} & \multicolumn{2}{c}{19.53 / 0.287 / 0.765} \\
        UTI~\cite{prior}   & \multicolumn{2}{c}{21.02 / 0.333 / 0.717} & \multicolumn{2}{c}{21.83 / 0.369 / 0.685} & \multicolumn{2}{c}{20.68 / 0.350 / 0.756} & \multicolumn{2}{c}{20.74 / 0.359 / 0.743} & \multicolumn{2}{c}{21.16 / 0.347 / 0.712} \\
        VIDUE~\cite{vidue}      & \multicolumn{2}{c}{21.63 / 0.356 / 0.721}  & \multicolumn{2}{c}{21.82 / 0.364 / 0.718} & \multicolumn{2}{c}{21.63 / 0.370 / 0.713} & \multicolumn{2}{c}{21.21 / 0.372 / 0.704} & \multicolumn{2}{c}{21.64 / 0.366 / 0.715} \\
        UEVD~\cite{uevd}+TL~\cite{timelens}& \multicolumn{2}{c}{24.26 / 0.813 / 0.433}         & \multicolumn{2}{c}{24.94 / 0.818 / 0.425} & \multicolumn{2}{c}{26.13 / 0.827 / 0.407} & \multicolumn{2}{c}{26.92 / 0.828 / 0.379} & \multicolumn{2}{c}{25.58 / 0.822 / 0.414} \\
        EVDI~\cite{evdi}        & \multicolumn{2}{c}{28.75 / 0.893 / \underline{0.324}}          & \multicolumn{2}{c}{28.62 / 0.887 / \underline{0.323}} & \multicolumn{2}{c}{27.09 / 0.870 / \underline{0.339}} & \multicolumn{2}{c}{24.16 / 0.786 / 0.386} & \multicolumn{2}{c}{27.33 / 0.875 / \underline{0.336}} \\
        REFID~\cite{refid}      & \multicolumn{2}{c}{\underline{30.51} / \underline{0.921} / 0.349}          & \multicolumn{2}{c}{\underline{30.91} / \underline{0.922} / 0.346} & \multicolumn{2}{c}{\underline{30.68} / \underline{0.918} / 0.346} & \multicolumn{2}{c}{\underline{29.59} / \underline{0.907} / \underline{0.342}} & \multicolumn{2}{c}{\underline{30.60} / \underline{0.919} / 0.346} \\
        EBFI~\cite{EBFI}  & \multicolumn{2}{c}{26.28 / 0.883 / 0.380}          & \multicolumn{2}{c}{26.18 / 0.878 / 0.37} & \multicolumn{2}{c}{25.24 / 0.860 / 0.350} & \multicolumn{2}{c}{23.95 / 0.829 / 0.382} & \multicolumn{2}{c}{25.54 / 0.866 / 0.355} \\
        \textbf{Ours}           & \multicolumn{2}{c}{\textbf{33.36 / 0.937 / 0.312}} & \multicolumn{2}{c}{\textbf{33.93 / 0.938 / 0.307}} & \multicolumn{2}{c}{\textbf{33.85 / 0.937 / 0.303}} & \multicolumn{2}{c}{\textbf{32.70 / 0.929 / 0.282}} & \multicolumn{2}{c}{\textbf{33.69 / 0.937 / 0.303}} \\ \hline
        \end{tabular}
        }
    
\label{tab:gopro_highrev_16sp}
\vspace{-6mm}
\end{table*}

\vspace{-6.5mm}
\section{Experiments}
\label{sec:experiments}
\vspace{-2mm}
\subsection{Experimental Setups}
\vspace{-2mm}
\label{ssec:setup}
\noindent \textbf{Datasets.}
We evaluate on GoPro~\cite{gopro} (with synthetic events from ESIM~\cite{esim}), HighREV~\cite{refid} (real events, $1632 \times 1224$), and RealBlur-DAVIS~\cite{EBFI} (real blurry videos and events).
To simulate motion blur in GoPro and HighREV, we average $m$ consecutive frames within the exposure duration and downsample them to shutter periods of 10 and 16 frames, denoted as 10$\Downarrow$ and 16$\Downarrow$.

\noindent \textbf{Implementation Details.}
Our framework is trained end-to-end without pre-training.
We train on GoPro for 300 epochs using AdamW~\cite{adamw} with a learning rate of \texttt{5e-4}, scheduled by cosine annealing~\cite{cosine_anneal}, and fine-tune on HighREV for 20 epochs with a learning rate of \texttt{1e-4}.
We follow the official data splits.
During training, $m$ is randomly sampled within the shutter period (e.g., $m \in [1, 9]$ for 10$\Downarrow$), and inputs are cropped to $128{\times}128$ (GoPro) or $256{\times}256$ (HighREV).
Target timestamps are randomly selected from available indices.

\noindent \textbf{Evaluation Protocols.}
We evaluate under two settings: (1) \textit{Symmetric exposure} with fixed $m$ values ($[1,5,9]$ for 10$\Downarrow$, $[1,5,11,15]$ for 16$\Downarrow$), and (2) \textit{RandEx}, where $m$ is randomly sampled as in training.
Metrics include PSNR (dB), SSIM~\cite{ssim}, and LPIPS~\cite{lpips}.
We additionally report temporal consistency metrics (tOF, tLP~\cite{TecoGAN}) and NIQE\cite{NIQE} for RealBlur-DAVIS, which lacks ground-truth frames.

\begin{figure}[t]
    \centering
    \includegraphics[clip, width=0.88\textwidth]{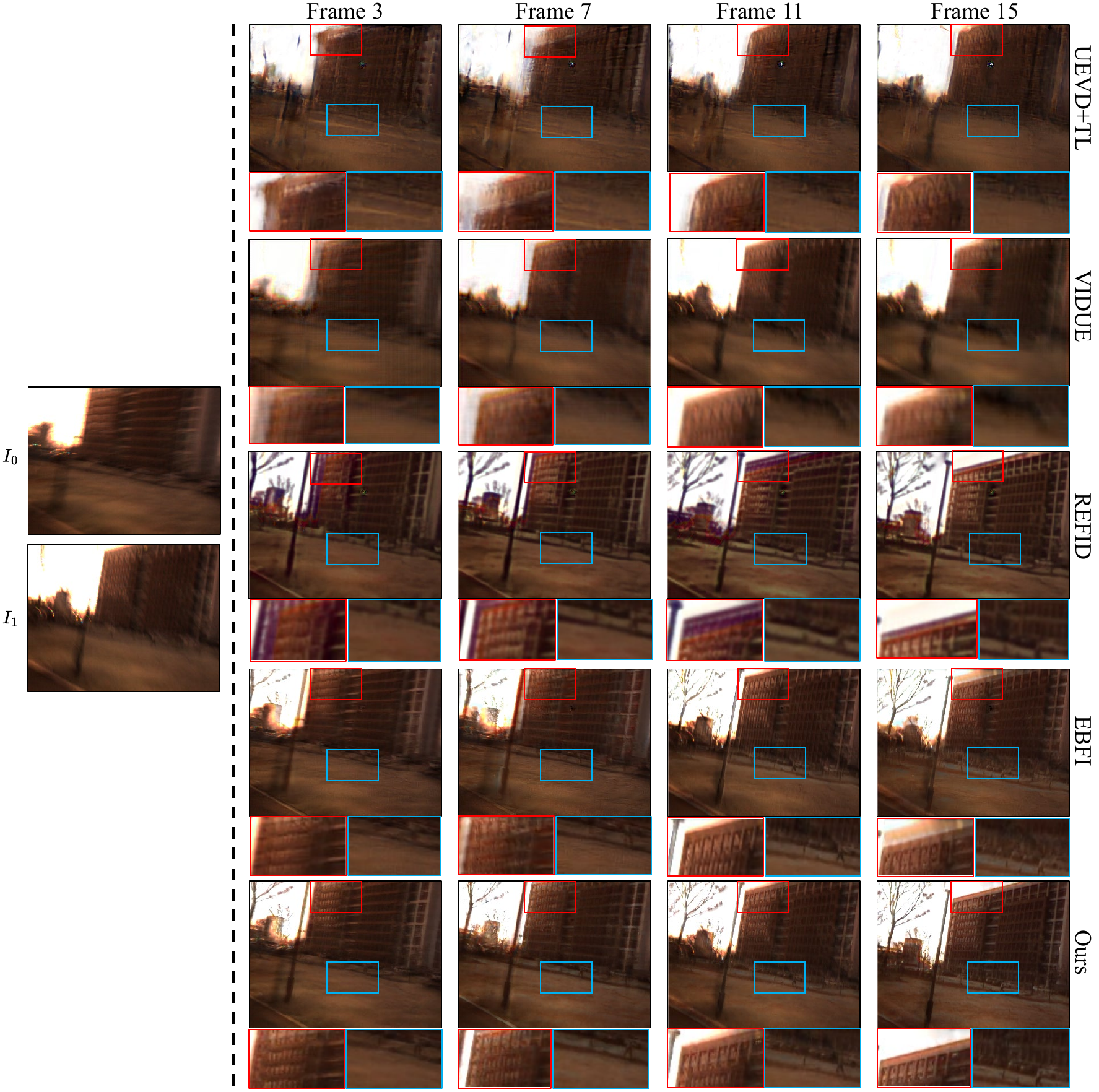}
    \vspace{-2mm}
    \caption{ 
        Qualitative results on RealBlur-DAVIS (real blur \& events)
    }
    \vspace{-6mm}
    \label{fig:realblur1}
\end{figure}

\vspace{-3mm}
\subsection{Comparison to the State-of-the-arts}
\vspace{-1mm}
\label{ssec:results}
Table~\ref{tab:summary} shows the summary of each compared method.
All methods are retrained for fair comparison, except TimeLens (TL), which is evaluated using its pre-trained model due to unavailable training code.
Exposure-specific baselines are provided with ground-truth exposure information.
All results are averaged across the restored frames.

\noindent \textbf{Quantitative Evaluation.} 
Tables~\ref{tab:gopro_highrev_10sp} and~\ref{tab:gopro_highrev_16sp} show results on GoPro and HighREV under 10$\Downarrow$ and 16$\Downarrow$ settings. 
As expected, RGB-only and cascaded methods tend to yield lower performance, likely due to motion ambiguity and error propagation, respectively.
Exposure-specific models (REFID, EVDI), trained with known exposure times, benefit from temporal alignment and produce comparable results.
EBFI, despite being exposure-agnostic, performs competitively on GoPro, but its performance declines on HighREV, where larger pixel displacement highlights the importance of temporal constraints.
Our method achieves the highest PSNR, SSIM, and LPIPS on both GoPro and HighREV, with margins up to 2.85dB over the second-best. Temporal consistency (tOF, tLP) and NIQE results on RealBlur-DAVIS are provided in the supplementary material (Sec.~\ref{ssec:more_quan}).

\noindent \textbf{Qualitative Evaluation.} 
We present qualitative results on RealBlur-DAVIS (real captured video) in Fig.~\ref{fig:realblur1}, where our method produces temporally consistent frames with minimal artifacts.
Additional comparisons on GoPro, HighREV (with synthetic events), and RealBlur-DAVIS, along with supplementary video clips demonstrating qualitative results and arbitrary temporal upscaling, are provided in supplementary material (Sec.~\ref{ssec:more_qual}).

\vspace{-5mm}
\subsection{Analysis}
\label{ssec:analysis}
\vspace{-2mm}


\begin{figure}[t]
    \centering
    \includegraphics[clip, width=0.55\linewidth]{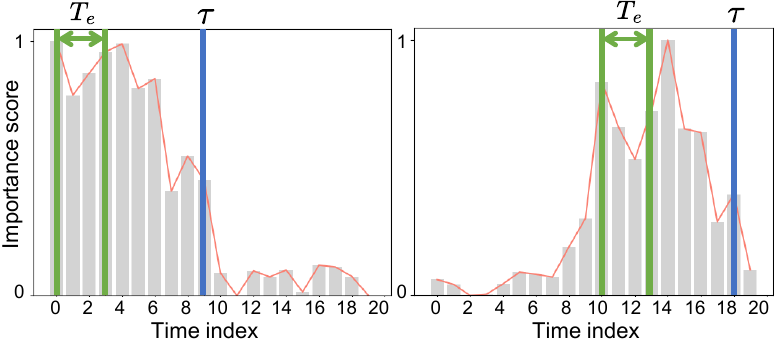}    
    \vspace{-2mm}
        \caption{
        Channel-wise importance score $\frac{\partial\mathcal{E}_{\tau \rightarrow [t_s, t_e]}}{\partial\mathbb{E}_N}$ indicating the sampling contribution of each time index. TES effectively selects events near the unknown exposure $T_e$ (\textcolor{Green}{green}) and target timestamp $\tau$ (\textcolor{MidnightBlue}{blue}), regardless of whether $\tau$ lies outside the exposure window.
        } 
    \vspace{-3mm}
    \label{fig:saliency_main} 
\end{figure}


\noindent\textbf{Analysis of the TES Module.}
We analyze whether TES effectively samples events around the target timestamp $\tau$ and the unknown exposure $T_e$.
To this end, we apply a gradient-based interpretation~\cite{gradcam} and compute importance scores as $\frac{\partial\mathcal{E}_{\tau \rightarrow T_e}}{\partial\mathbb{E}_N}$, indicating the contribution of each event to the sampled output.
Given that input events are divided into $N$ temporal bins and stacked channel-wise, we report the average importance score per channel on GoPro-10$\Downarrow$.
As shown in Fig.~\ref{fig:saliency_main}, TES consistently focuses on events near both $T_e$ and $\tau$, even when $\tau$ lies outside the exposure window.

We present ablation results on GoPro-10$\Downarrow$ (RandEx) in the `TES' section of Table~\ref{tab:ablation}.
We evaluate the impact of removing key components from the TES module: (1) the sampling process in Eq.~(\ref{eq:corr_score}) and (2) the positional encoding (PE) of the target timestamp $\tau$.
Removing the sampling process hinders the TES module from focusing on events near $\tau$ and the blind exposure $T_e$, while excluding PE causes the module to ignore $\tau$ and focus only on $T_e$.
These degradations confirm that both components are essential for effective event sampling. Further qualitative ablations are included in the supplementary (Sec.~\ref{ssec:more_analysis}).




\begin{figure}[!t]
    \centering
    \includegraphics[clip, width=0.93\textwidth]{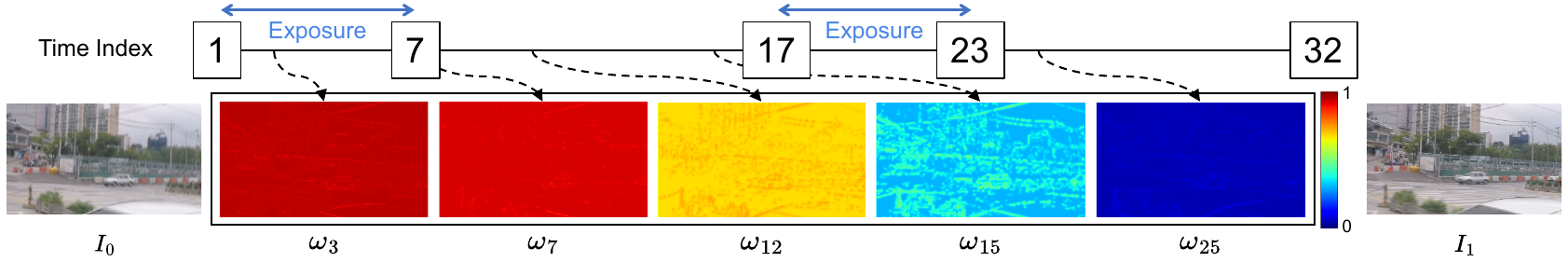}
    \vspace{-3mm}
    \caption{ 
        Visualization of importance maps $\omega_\tau$ for $\tau \in {3, 7, 12, 15, 25}$, with symmetric exposure durations ($m=7$) for $I_0$ and $I_1$.
    }
    \vspace{-5mm}
    \label{fig:immap}
\end{figure}

\begin{wraptable}{r}{0.45\textwidth}
\vspace{-2mm}
\caption{Ablation study on GoPro-10$\downarrow$ (RandEx). All ablations are retrained from scratch, except Swap $\omega_\tau$.}
\centering
\small
\resizebox{0.45\textwidth}{!}{
\setlength{\tabcolsep}{10pt}
\renewcommand\arraystretch{1.0}
\begin{tabular}{ l c c  c  c}
    \bottomrule[0.15em]
    \rowcolor{tableHeadGray}
    \textbf{Ablation}                     &                & \textbf{PSNR} $\uparrow$ & \textbf{SSIM} $\uparrow$ & \textbf{LPIPS} $\downarrow$ \\ \hline \hline
    \textbf{TES} (Fig.~\ref{fig:tes}) \\ 
    \hline
    \multicolumn{2}{l}{Remove Sampling}            & 32.05   & 0.948  & 0.058  \\
    \multicolumn{2}{l}{Remove $\tau$}              & 31.12   & 0.937 & 0.073  \\  
    \hline
    \textbf{TIM} (Fig.~\ref{fig:tim}) \\ 
    \hline
    \multicolumn{2}{l}{Remove $\tau$}              & \multicolumn{3}{c}{\womark}                 \\
    \multicolumn{2}{l}{Remove Eq.~\ref{eq:tim_step1}}    & 32.24 & 0.945  & 0.051 \\ 
    \multicolumn{2}{l}{Fixed $\omega_\tau=0.5$}    & 31.51  & 0.939   & 0.068  \\
    \multicolumn{2}{l}{Swap $\omega_\tau$}         & 20.43  & 0.588   & 0.194  \\
    \hline
    \multicolumn{2}{l}{\textbf{Ours}}              & \textbf{33.39} & \textbf{0.961} & \textbf{0.045} \\ \hline
\end{tabular}
}
\vspace{-2mm}
\label{tab:ablation}
\end{wraptable}

\noindent\textbf{Analysis of the TIM Module.}
We visualize the importance map $\omega_\tau$ in Fig.~\ref{fig:immap} using GoPro-16$\Downarrow$ data with $m=7$, where the exposure ranges are $[1, 7]$ for $I_0$ and $[17, 23]$ for $I_1$.
As $\tau$ approaches the exposure of $I_0$, $\omega_\tau$ assigns greater weights to features from $I_0$, and likewise shifts toward $I_1$ as $\tau$ nears its exposure.
These results show that TIM adaptively emphasizes features that are temporally closer and spatially relevant to the target timestamp.

We report ablation results for the TIM module in the `TIM' section of Table~\ref{tab:ablation}.
Removing the target timestamp $\tau$ causes convergence failure, underscoring its importance for generating $\omega_\tau$.
Disabling the attention mechanism in Eq.~\ref{eq:tim_step1} also degrades performance, confirming the benefit of spatial relevance.
We also evaluate two variants: replacing $\omega_\tau$ with a constant value (0.5) and swapping its order during feature blending in Eq.~(\ref{eq:TFB}).
Both degrade performance, with the swapped version performing notably worse, validating the importance and learned structure of $\omega_\tau$.
Qualitative ablations are provided in the supplementary (Sec.~\ref{ssec:more_analysis}).


\vspace{-3mm}
\section{Conclusion}
\vspace{-3mm}
In this work, we proposed a simple yet effective framework for event-guided, exposure-agnostic video frame interpolation that explicitly incorporates temporal constraints via adaptive feature blending. Our method integrates two synergistic modules: the Target-adaptive Event Sampling module for event sampling around the target timestamp and unknown exposure, and the Target-adaptive Importance Mapping module for weighting features based on temporal and spatial relevance. Extensive experiments validate its effectiveness across synthetic and real-world datasets.

\vspace{-3mm}
\section*{Acknowledgements}
\vspace{-3mm}
This work was supported by Institute of Information \& communications Technology Planning \& Evaluation (IITP) grant funded by the Korea government (MSIT) (No.RS-2025-25443318, Physically-grounded Intelligence: A Dual Competency Approach to Embodied AGI through Constructing and Reasoning in the Real World).
This work was supported by Institute of Information \& communications Technology Planning \& Evaluation(IITP) grant funded by the Korea government (MSIT) (RS-2023-00237965, Recognition, Action and Interaction Algorithms for Open-world Robot Service).
This work was supported by the National Research Foundation of Korea (NRF) grant funded by the Korea government (MSIT) (No. RS-2023-00208506).
This work was supported in part by the Samsung Electronics Company Ltd, System LSI Division (IO201210-07984-01).



\bibliography{egbib}
\clearpage
\newcommand{\appendixnumbering}{
    \renewcommand{\thefigure}{S\arabic{figure}}
    \renewcommand{\thetable}{S\arabic{table}}
}

\appendix
\appendixnumbering

\setcounter{page}{1}
\setcounter{table}{0}
\setcounter{figure}{0}

\begin{table}[t]
\caption{tOF$\downarrow$ / tLP$\downarrow$ evaluations on GoPro and HighREV with RandEX setting.}
\vspace{2mm}
\centering
\small
\resizebox{1.0\textwidth}{!}{
\setlength{\tabcolsep}{10pt}
\renewcommand\arraystretch{1.0}
    \begin{tabular}{l c c c c c c c c}
        \bottomrule[0.15em]
        \rowcolor{tableHeadGray}
                 \multicolumn{9}{c}{\textbf{tOF$\downarrow$ / tLP$\downarrow$}}     \\ \hline \hline
        \textbf{Dataset}  &Ours  & EBFI & REFID & EVDI & UEVD + TL & VIDUE & UTI & NAFnet + RIFE \\ \hline
        GoPro-$10\Downarrow$  & \textbf{0.0067} / \textbf{0.6366} & 0.0082 / 0.8873 & 0.0078 / 0.7618 & 0.0104 / 0.8324 & 0.0206 / 1.4478 & 0.0232 / 1.5676 & 0.0410 / 3.9124 & 0.0495 / 4.6272 \\
        GoPro-$16\Downarrow$  & \textbf{0.0867} / \textbf{0.7618}  & 0.1064 / 0.9227 & 0.1073 / 0.9827 & 0.1270 / 1.0581 & 0.469 / 6.2587 & 0.337 / 2.8956 & 0.398 / 4.9081 & 0.404 / 5.0532 \\
        \hline
        HighREV-$10\Downarrow$ & \textbf{0.0048} / \textbf{4.212} & 0.0064 / 6.522 & 0.0052 / 5.691 & 0.0052 / 5.752 & 0.0094 / 7.231 & 0.0157 / 8.078 & 0.0162 / 11.814 & 0.0158 / 10.584 \\
        HighREV-$16\Downarrow$ & \textbf{0.0053} / \textbf{4.556} & 0.0083 / 7.145 & 0.0059 / 6.170 & 0.0060 / 6.132 & 0.0114 / 7.543 & 0.0158 / 9.072 & 0.0173 / 11.648 & 0.0177 / 10.329 \\
        \hline
    \end{tabular}
}
\label{tab:temp_consistency}
\end{table}

\begin{table}[t]
\caption{NIQE evaluations on RealBlur-DAVIS.}
\vspace{2mm}
\centering
\small
\resizebox{0.9\textwidth}{!}{
\setlength{\tabcolsep}{10pt}
\renewcommand\arraystretch{1.0}
    \begin{tabular}{l c c c c c c c c}
        \bottomrule[0.15em]
        \rowcolor{tableHeadGray}
                 \multicolumn{9}{c}{\textbf{NIQE$\downarrow$}}     \\ \hline \hline
        \textbf{Dataset}  &Ours  & EBFI & REFID & EVDI & UEVD + TL & VIDUE & UTI & NAFnet + RIFE \\ \hline
        RealBlur-DAVIS  & \textbf{26.55} & 28.62 & 29.48 & 38.97 & 37.79 & 36.72 & 40.67 & 41.14 \\
        \hline

    \end{tabular}
}
\label{tab:niqe}
\end{table}

In this supplementary material, we first present additional results in Sec.~\ref{sec:more_experiments}.
Next, we provide network details in Sec.~\ref{sec:detail}.
Finally, Sec.~\ref{sec:limitation} discusses the limitations of our approach.

\begin{table}[t]
\caption{Computational cost on GoPro-10$\Downarrow$ (RandEx setting). Runtime measured on a TITAN RTX GPU.}
\vspace{2mm}
\centering
\small
\resizebox{1.0\linewidth}{!}{
\setlength{\tabcolsep}{10pt}
\renewcommand\arraystretch{1.0}
    \begin{tabular}{l c c c c c c c c}
        \bottomrule[0.15em]
        \rowcolor{tableHeadGray}
        \textbf{Complexity}    &Ours  &REFID &EBFI  & EVDI &UEVD + TL & VIDUE & UTI & NAFnet + RIFE\\ \hline
        \# Params (M) & 10.58 & 15.76   & \underline{4.23} & \textbf{0.396} & 107.09 & 62.46 & 49.02 & 70.07 \\
        FLOPs (G) & 166.83  & 182.04 & \underline{70.08} & \textbf{53.08} & 671.83 & 285.93 & 221.09 & 189.66 \\
        Runtime (ms) & \underline{90.93}  & 164.6   & 96.23 & \textbf{69.2} & 420.4 & 195.4 & 192.3 & 152.4\\
        \hline
        PSNR      &   \textbf{33.39}    &   \underline{31.03}   &  30.89  &  27.51 & 23.51 &  25.86 & 22.17 & 21.05\\
        \hline

    \end{tabular}
}
\label{tab:complexity}
\end{table}

\begin{figure}[t]
    \centering
    \subfigure[]{%
        \includegraphics[width=0.35\textwidth]{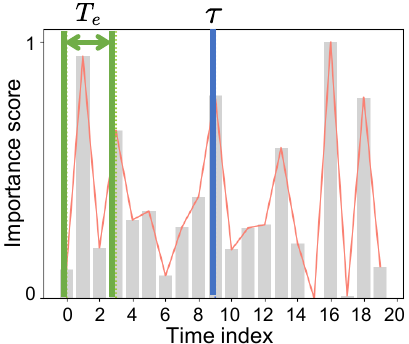}
        \label{fig:saliency_sub2}
    }
    \subfigure[]{%
        \includegraphics[width=0.35\textwidth]{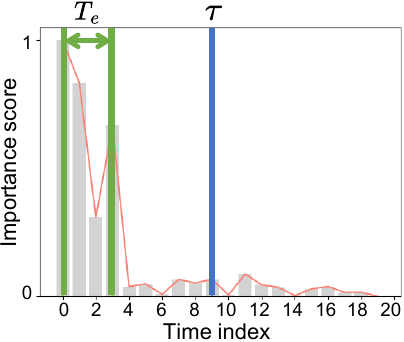}
        \label{fig:saliency_sub3}
    }
    \hfill
    \caption{
        Channel-wise average importance scores $\frac{\partial\mathcal{E}_{\tau \rightarrow [t_s, t_e]}}{\partial\mathbb{E}_N}$ showing how much each temporal slice in $\mathbb{E}_N$ contributes to the sampled events.
        (a) Without the sampling process and (b) without the target timestamp $\tau$. 
        Higher scores indicate stronger selection by the TES module.
    }
    \label{fig:supple_saliency}
\end{figure}

\section{Additional Results}
\label{sec:more_experiments}
\subsection{More Quantitative Results}
\label{ssec:more_quan}
We additionally report temporal consistency metrics (tOF, tLP~\cite{TecoGAN}) and the no-reference quality metric NIQE~\cite{NIQE} for RealBlur-DAVIS, which lacks ground-truth frames.
As shown in Table~\ref{tab:temp_consistency}, our method achieves the most temporally coherent results on both GoPro and HighREV datasets.
Furthermore, Table~\ref{tab:niqe} shows that our method also achieves the best NIQE scores, indicating superior perceptual quality on real-world blurry videos.

\subsection{More Qualitative Results}
\label{ssec:more_qual}
\begin{figure}[!t] 
    \centering
    \includegraphics[clip, width=0.8\linewidth]{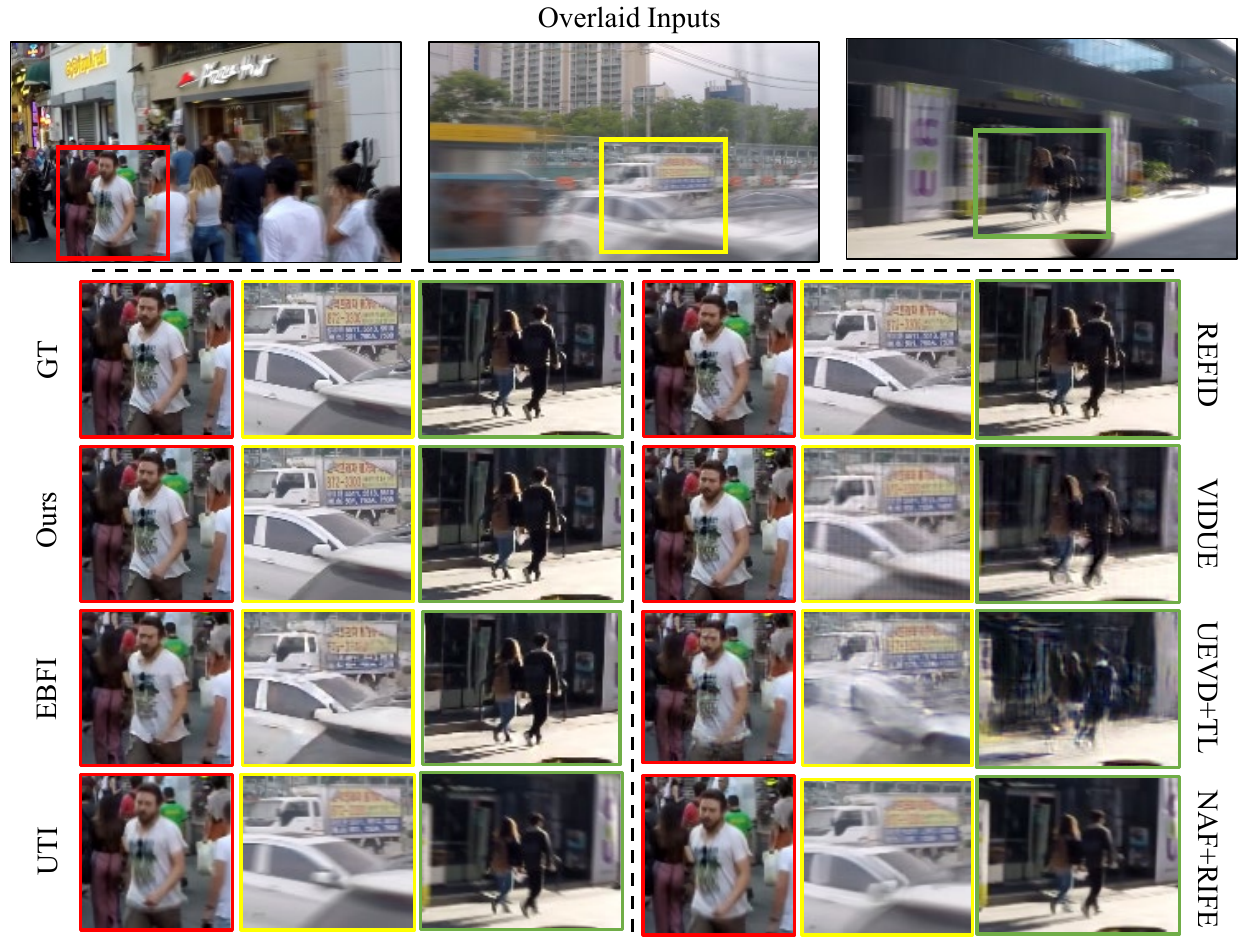}
    \vspace{-3mm}
    \caption{ 
      Qualitative results on GoPro (synthetic blur \& events). 
    }
    \vspace{-3mm}
    \label{fig:gopro16}
\end{figure}

\begin{figure}[t]
    \centering
    \includegraphics[width=0.9\textwidth]{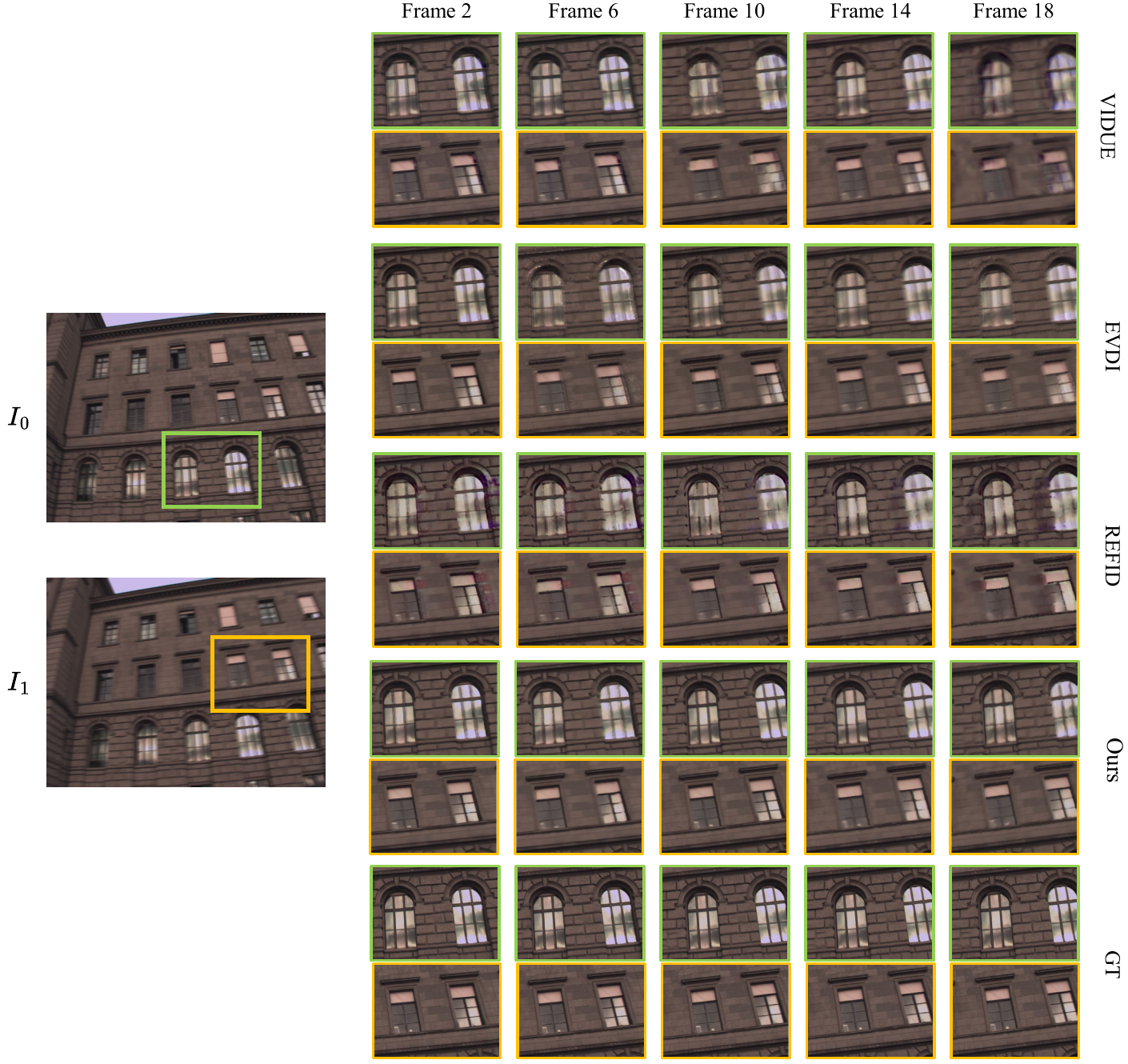}
    \vspace{-4mm}
    \caption{ 
        Qualitative results on HighREV (synthetic blur \& real events).
       }
    \label{fig:highrev}
\end{figure}

\begin{figure}[t]
    \centering
    \includegraphics[width=0.76\textwidth]{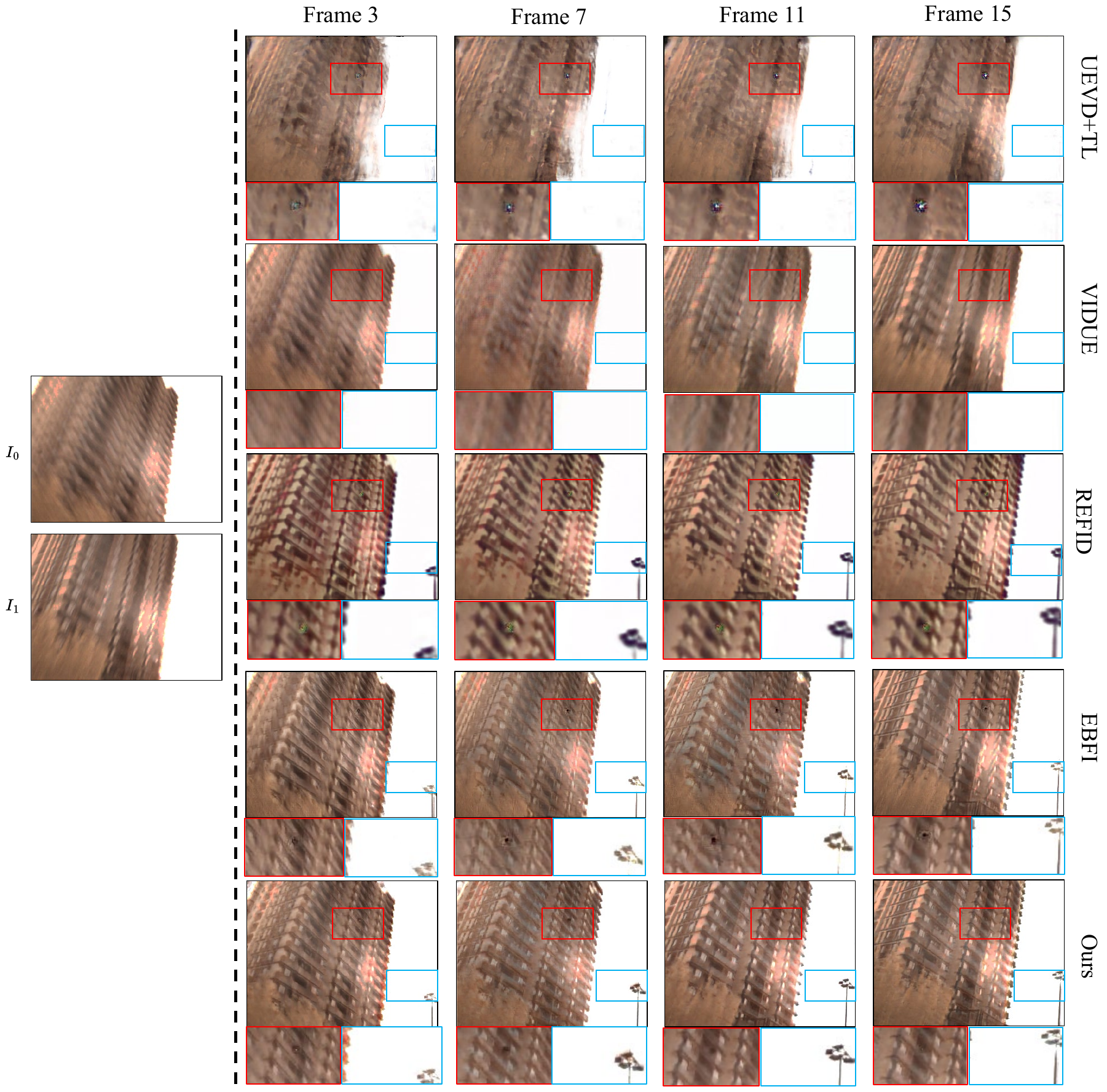}
    \vspace{-4mm}
    \caption{ 
       More qualitative results on RealBlur-DAVIS (real blur \& events)
    }
    \vspace{-3mm}
    \label{fig:realblur2}
\end{figure}

\begin{figure}[t]
    \centering
    \includegraphics[width=0.76\textwidth]{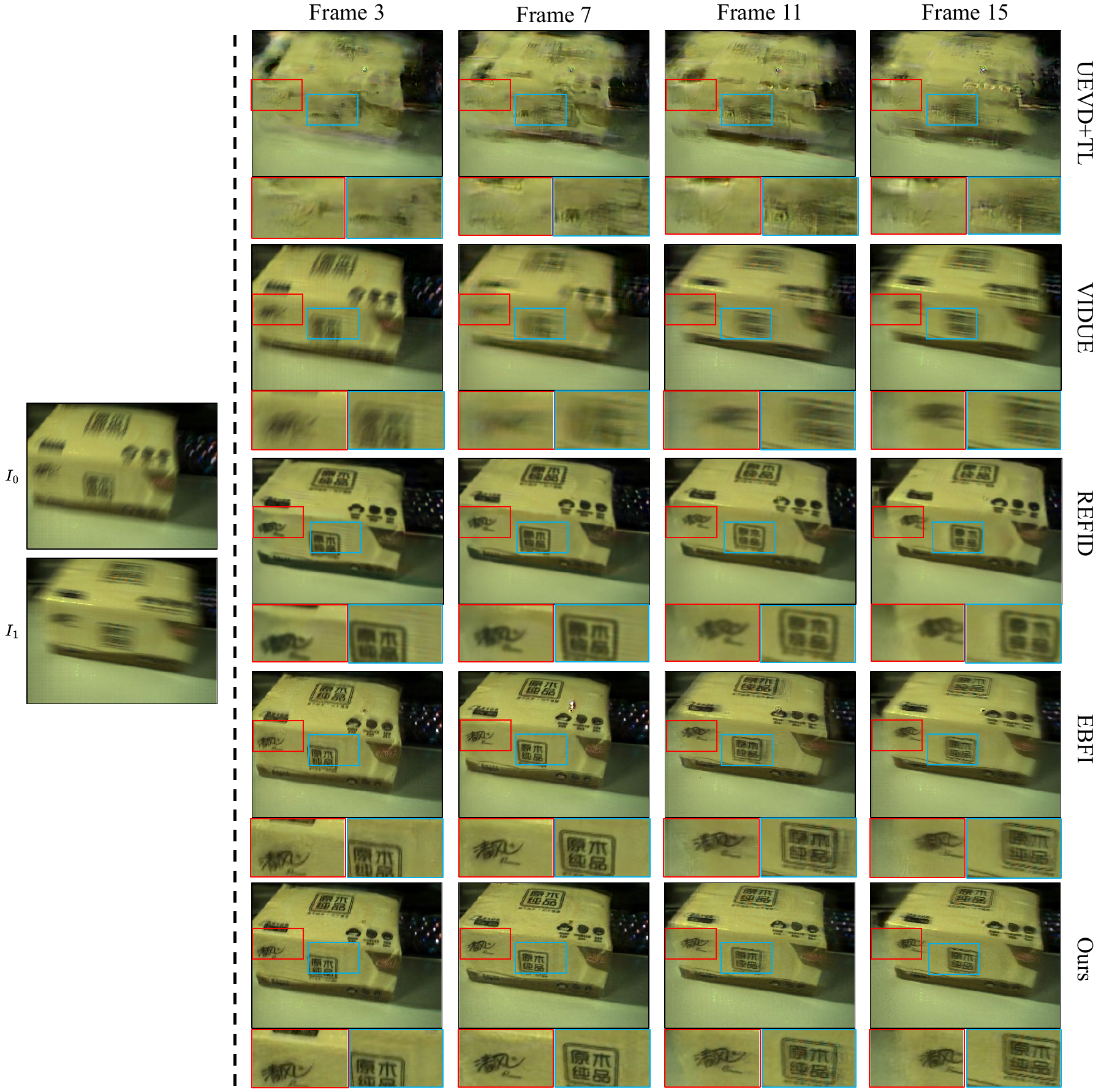}
    \vspace{-4mm}
    \caption{ 
       (continued) More qualitative results on RealBlur-DAVIS. 
    }
    \label{fig:realblur3}
\end{figure}

We present additional visual comparisons on GoPro, HighREV, and RealBlur-DAVIS in Fig.\ref{fig:gopro16}, Fig.\ref{fig:highrev}, Fig.\ref{fig:realblur2}, and Fig.\ref{fig:realblur3}.
On the synthetic GoPro dataset (Fig.\ref{fig:gopro16}), our method consistently produces the most faithful results across diverse scenes.
For HighREV (Fig.\ref{fig:highrev}), visualizations across various target timestamps show that our method maintains high-quality reconstructions with minimal artifacts.

On the RealBlur-DAVIS dataset (Figs.\ref{fig:realblur2} and\ref{fig:realblur3}), two-stage methods (UEVD + TL) yield the least effective results, while event-guided approaches (Ours, EBFI, REFID) outperform the RGB-only baseline (VIDUE) by leveraging precise motion cues from events.
Notably, as seen in Fig.~\ref{fig:realblur2}, event-guided methods can even recover objects missing from RGB frames thanks to the high dynamic range of event sensors.
Among them, our method delivers the most reliable results across all target timestamps.

We also provide two types of supplementary MP4 video clips: (1) comparisons with strong baselines (EBFI~\cite{EBFI} and REFID~\cite{refid}), and (2) demonstrations showing our method upscaling low-frame-rate blurry inputs into sharp, high-frame-rate outputs at arbitrary temporal scales.
These videos further highlight the temporal coherence and scalability of our approach.

\begin{figure}[t]
    \centering
    \includegraphics[width=0.35\textwidth]{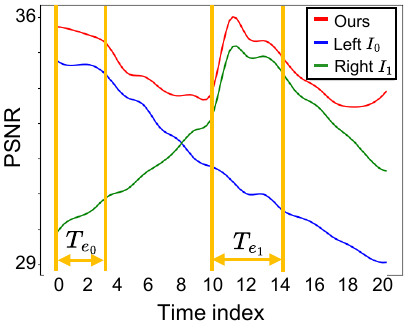}
    \caption{ 
        PSNR comparison over time with and without temporal constraints.
        Our method (\textcolor{Red}{red}) maintains stable performance across target timestamps, while the version without temporal constraints (\textcolor{blue}{blue} \& \textcolor{ForestGreen}{green}) shows degradation.
       }
    \label{fig:psnr}
\end{figure}

\subsection{More Analysis}
\label{ssec:more_analysis}
We qualitatively ablate the TES module by removing the sampling step and the positional encoding of the target timestamp $\tau$.
Figure~\ref{fig:supple_saliency} illustrates the resulting changes in channel-wise importance scores.
Without sampling (Fig.\ref{fig:supple_saliency}(a)), TES fails to focus on events around $\tau$ and $T_e$; without $\tau$ (Fig.\ref{fig:supple_saliency}(b)), it only attends to $T_e$.
These results highlight the critical role of both components in effective event sampling.

To further evaluate the effectiveness of the TIM module, Fig.~\ref{fig:psnr} compares the PSNR over time between our full model and a variant that uses only a single frame and its corresponding events.
The results underscore the role of TIM in enhancing temporal consistency across frames.

\noindent \textbf{Computational Cost.} 
Table~\ref{tab:complexity} compares model parameters, FLOPs, and runtime.
EVDI~\cite{evdi} has the lowest cost due to its lightweight design, but shows limited performance under unknown and dynamic exposures.
EBFI~\cite{EBFI} is more efficient than ours in parameters and FLOPs, but lacks temporal modeling, leading to lower performance.
Our method, while relatively heavier in parameters and FLOPs, achieves substantially better performance with comparable runtime by leveraging temporal constraints.

\vspace{-4mm}
\section{Network Details}
\label{sec:detail}
\begin{figure}[t]
    \centering
    \includegraphics[clip, width=0.6\textwidth]{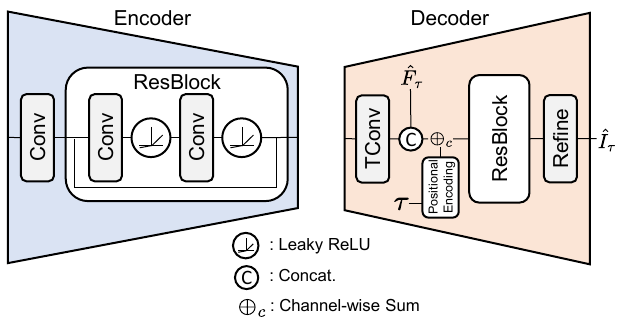}
    \caption{
        Detailed architecture of the encoder and decoder.
        Frame and event encoders share the same structure. The model follows a U-Net design with three encoders and two decoders, where only the final decoder includes a refinement block.
    }
    \vspace{-3mm}
    \label{fig:encdec}
\end{figure}

\vspace{-2mm}
Fig.~\ref{fig:encdec} illustrates the detailed architectures of the encoder and decoder, which are not fully described in the main paper.
The frame encoder processes the consecutive captured frames, while the event encoder extracts features from the stacked events produced by the TES module.
These features are then fused via a channel-wise attention mechanism~\cite{efnet}.
The fused representations are adaptively blended by the TIM module to produce the target feature $\hat{F}_\tau$.
The decoder takes $\hat{F}_\tau$ along with the positional encoding of the target timestamp $\tau$ to reconstruct the sharp target frame $\hat{I}_\tau$.
A final refinement step is applied using the fusion network adopted from~\cite{evdi}.

\vspace{-4mm}
\section{Limitation and Future Work} 
\label{sec:limitation}
\vspace{-2mm}
Although we have shown the effectiveness of our method, it has certain limitations that are worth further exploration.
Specifically, the current framework assumes that frames and events share the same spatial resolution. 
In real-world scenarios, however, frame-based cameras and event sensors often operate at different resolutions~\cite{event_vision_survey}.
To enhance the practicality of our approach, future work could explore strategies for fusing frame and event data with mismatched resolutions.
One promising direction is to incorporate implicit neural representations for resolution-agnostic fusion~\cite{implicit, gem}, which we leave as a potential avenue for future research.


\end{document}